# Wavelength-selective bidirectional plasticity in self-powered Se/ZnMgO visual synapses

Zacharie Jehl Li-Kao[1,*], Miho Akasaka[2], Myeongok Kim[3], Kunal J. Tiwari[4], Antoine Moore[5], Mykhailo A. Koltsov[6], Sergio Giraldo[1], Saiki Kitagawa[2], Marcel Placidi[1], Taizo Kobayashi[2,**]

[1] Electronic Engineering Department, Photovoltaic Lab – Micro and Nano Technologies Group (MNT), Universitat Politècnica de Catalunya (UPC), EEBE, Av. Eduard Maristany 10-14, 08019 Barcelona, Spain

[2] Department of Mechanical Engineering, Ritsumeikan University, 1-1-1 Nojihigashi, Kusatsu, Shiga 525-8577, Japan

[3] Research Center for Advanced Science and Technology, The University of Tokyo, 4-6-1 Komaba, Meguro-ku, Tokyo, Japan

[4] Department of Dielectrics, Institute of Physics of the Czech Academy of Sciences, 182 00 Prague 8, Czech Republic

[5] Stephenson Institute for Renewable Energy, Department of Physics, University of Liverpool, Liverpool, UK

[6] Department of Materials and Environmental Technology, Tallinn University of Technology, Ehitajate tee 5, Tallinn, 19086, Estonia

[*] Correspondence: zacharie.jehl@upc.edu

[**] Correspondence: tkt07004@fc.ritsumei.ac.jp

## Abstract

Artificial visual synapses combine sensing, memory and processing in a single device, but most of them need an external voltage to read the synaptic weight and light is generally able to strengthen the weight rather than to weaken it. We report here Se/ZnMgO thin film photodiodes in which the photocurrent at zero bias writes and reads the synaptic weight, with a direction of change set by the photon energy alone. At equal photon flux, 365 nm pulses increase the photocurrent by 38 % while 455 nm pulses decrease it, and the depression reaching 30 % at higher flux, with up to 19 distinguishable levels per branch and no electrical energy spent. Spectrally resolved light soaking shows that short wavelengths potentiate while low-energy light erases. The interesting property is therefore the photon energy-dependent competition between potentiation and depression, the crossover being set by the defect population of the interface. Optical simulations show that both wavelengths are absorbed within the first tens of nanometres of Se, and the stored state is erased by a reverse bias in a voltage-dependent way and by sub-gap light, which indicates that potentiation corresponds to interface defects filled with electrons, in agreement with drift-diffusion simulations. The devices also show relearning, their response is modulated by bending on flexible substrates, and a network built from the measured weight updates reaches 89 % accuracy on handwritten digit recognition without noise and 85 % with the measured noise, against 97.5 % for floating-point weights.

## 1. Introduction

The separation of processing and memory in the von Neumann architecture leads to the well known bottleneck which becomes critical for tasks such as image recognition and real-time sensory processing [1]. Bringing memory and processing in a single physical device has emerged as a promising route to address this. Artificial synapses, where conductance changes in response to an external stimulus, can be viewed as the basic building blocks of such systems. The idea is to mimic the plasticity of biological synapses where repeated stimulation strengthens or weakens the connection between two neurons [2].

Amongst the stimuli used to drive artificial synapses, light is particularly important. Visual synapses use photons as the input signal and they combine sensing, memory and processing in a single device, for in-sensor neuromorphic computing [3]. A wide range of material systems has been explored in this context, including oxide thin films [4], two-dimensional materials [5,6], organic semiconductors [7,8] and halide perovskites [9]. In most cases the synaptic weight is read by applying a voltage and measuring the persistent photoconductance after illumination. This reading step introduces an additional power input which reduces the energy efficiency of the system.

Self-powered synapses remove the need for a read voltage by monitoring the photocurrent in short circuit conditions instead, therefore the light stimulus simultaneously writes and reads the synaptic weight. This approach is rooted in photovoltaics: the photocurrent of a photodiode at zero bias depends on the internal state of the device, and it can be modified by previous illumination through the well-known light soaking effect [10], which is usually seen as detrimental to PV performance. Zero-bias synaptic operation has now been demonstrated in several material systems, including silicon-based devices [11], organic asymmetric heterojunctions [12] and lead-free perovskite stacks [13,14]. In most of these devices however, light is only able to potentiate the synaptic weight, and depression relies on an electrical pulse or on a slow relaxation in the dark; in a recent $CsPbBr_3$-based device, erasing is also optical but is selected by the direction of illumination rather than by its spectrum [15]. All-optical bidirectional plasticity where one wavelength potentiates and another one depresses the weight, has been reported in devices operating under an applied bias [16–19], but its combination with zero-bias operation remains largely unexplored to our knowledge, and the few existing reports are on a photoelectrochemical cell in aqueous electrolyte [20] or on the competition between photovoltaic and photothermoelectric effects in a hybrid heterojunction [21]. Such a property is attractive due to the fact that information can then be encoded in the spectral content of the optical signal, which is a feature that biological retinas already have through the wavelength selectivity of the cone photoreceptors.

In this work, we show that Se/ZnMgO thin film photodiodes, which were originally investigated for photovoltaics and for their piezophototronic response [22], can behave as self-powered visual synapses whose weight is increased or decreased depending on the photon energy only. We first establish the bidirectional plasticity at zero bias and equal photon flux, and how it is tuned by the photon dose. Spectrally resolved light soaking then shows that the sign of the change is set by a crossover photon energy, and optical simulations of the actual stack allow to separate the role of the photon energy from the role of the absorption profile. The stored state is erased by a reverse bias in a voltage-dependent way, which is consistent with a field-controlled charge-state population of interface defects, and the devices show relearning and keep operating on flexible substrates. Finally, the measured weight updates are used in a network simulation in order to assess the device for in-sensor computing.

## 2. Experimental section

**Device fabrication.** Se/ZnMgO photodiodes were fabricated on soda lime glass (SLG) and on 125 µm poly(ethylene terephthalate) (PET) substrates (product number 0081, GEOMATEC Co., Ltd, Japan). The device stack consists of ITO (150 nm) / ZnMgO (300 nm) / Se (500 nm) / $MoO_3$ (20 nm) / Au (80 nm), with an active area of 8 $mm^2$. The ZnMgO layer was deposited by sputtering and the Se, $MoO_3$ and Au layers by thermal evaporation, followed by annealing; the full fabrication details are given in Ref. [22]. Two fabrication batches with the same stack were characterised: batch A was measured in December 2024 (zero-bias plasticity at 365 and 455 nm) and batch B in January 2024 (dark conductance, spectral light soaking, electrical reset, relearning, flexible devices, and zero-bias reset).

**Self-powered synaptic measurements.** Light pulses were generated with Thorlabs fibre-coupled LEDs at 365 nm and 455 nm driven by a waveform generator controlling the pulse frequency, duration and number. The maximum photon flux available at the sample was $3.1 \times 10^{14}$ photons $cm^{-2}$ $s^{-1}$ at 365 nm (0.17 mW cm$^{-2}$) and 1.5 × 10$^{15}$ photons cm$^{-2}$ s$^{-1}$ at 455 nm (0.66 mW cm$^{-2}$). For the comparison of the two wavelengths, both fluxes were set to 3.2 × 10$^{14}$ photons cm$^{-2}$ s$^{-1}$. The photocurrent in short circuit (0 V) was recorded during each light pulse with a Keysight 34461A digital multimeter triggered by the waveform generator, with 3 to 5 samples per pulse. The value assigned to each pulse is the mean of the samples taken while the LED is on; samples below 50 % of the local bright level (LED rising edge, or sample taken after the LED switched off) are excluded, spurious triggers recording a dark block are discarded, and blocks whose samples were shifted in time with respect to the LED pulse, so that they fall on its rising edge or tail rather than on the plateau as in the rest of the run, or deviating by more than four robust standard deviations from the running median, are shown as open symbols and excluded from fits and plateau statistics. The first two pulses of each train are treated in the same way: at the start of a train, the timing of the trigger with respect to the LED pulse was not stable and the first pulse after a rest in the dark rises more slowly, so that their values cannot be compared with the rest of the train. Relative changes are therefore given with respect to the mean of pulses 3 to 5. The same rule is applied to every run. Runs belonging to one series were acquired in sequence, with rests of one to a few minutes in the dark between them.

**Dark conductance measurements.** For batch B, the persistent photoconductance was monitored by recording the current under a read voltage (0.5 V for the reset and relearning experiments, about 0.7 V for the flexible device) while 455 nm light pulses were applied, and the conductance was calculated as $G = I/V_{\text{read}}$. The stability of the current under the read voltage alone is shown in Figure S1: over 548 s, the standard deviation is 0.9 % of the mean and the drift is 0.06 % per minute. For the spectral light soaking, current-voltage curves were measured in the dark before and after successive 5 min exposures to LEDs at 1450, 850, 660, 530, 505, 455 and 365 nm, in this order, followed by 1450 nm exposures.

**Flexible devices.** Devices on PET were bent with the bending stage described in Ref. [22], using displacements of 2.5 mm in compression and in tension and a larger displacement in tension. Each bending state was measured once with trains of 100 pulses of 5 ms at 2 to 160 Hz and with a train of 1000 pulses at 160 Hz, and the flat state was measured again after the bending sequence.

**Optical and device simulations.** The absorption in each layer and the generation profile were calculated with the transfer-matrix method of the SOLEY simulation platform [23], using measured optical constants for every layer: ZnMgO from spectroscopic ellipsometry of $Mg_xZn_{1-x}O$ with $x = 0$, 0.12, 0.26 and 0.38 [24] (the Mg fraction of the sputtered layer is not known precisely and is therefore bracketed), trigonal Se, ITO, $MoO_x$ and Au. The LED lines were described by Gaussians of 9 and 18 nm full width at half maximum. Drift-diffusion simulations were performed with SCAPS-1D [25], with the parameters given in Table S1.

**Network simulations.** The measured potentiation and depression curves were fitted with the soft-bound update of the SYNAPSYS simulator [26], $\Delta G = A(G_{\max} - G)^{\alpha}$ for potentiation and $\Delta G = -B(G - G_{\min})^{\beta}$ for depression, and the nonlinearity was quantified with the NeuroSim parameter $\nu$ [27]. A 784-100-10 perceptron was trained on MNIST [28], with each weight being stored as a differential pair of devices updated only through the fitted potentiation (365 nm) and depression (455 nm) pulses, with a finite number of levels and with the measured cycle-to-cycle scatter, and compared with floating-point training (ten seeds per condition, and twenty when the measured cycle-to-cycle scatter is included, since this case varies more from one run to another). The nonlinearity $\nu$ is defined on the curve fitted to the pulses 3 to 60 of each train, the state of each branch being normalised to its own measured window, and the values obtained directly from the measured points are the same within 1 % (0.144 and 0.632 for the potentiation and the depression, respectively). The details of the network and of the pulse updates and of the noise are given in Section S6.

## 3. Results and discussion

### 3.1. Bidirectional plasticity at zero bias

The device architecture is shown on Figure 1a. Light soaking is a well-documented phenomenon in thin film photovoltaics and is generally attributed to metastable defects at interfaces and in the bulk [10,29]. While this effect is usually seen as detrimental in PV, it can be used for simulating visual memory, by taking advantage of the persistent modification of the photocurrent similarly to the change of synaptic weight after stimulation.

In order to characterise the self-powered behaviour, we stimulate the photodiodes with trains of light pulses at zero bias and record the photocurrent of each pulse. THe device operates without any electrical bias so the incident light pulse is at the same time the write signal, which modifies the state of the device, and the read signal, the photocurrent reflecting that state. Figure 1b compares trains of 60 pulses of 5 ms at 20 Hz under 365 nm and 455 nm light at the same photon flux, recorded three minutes apart on the same device. At the beginning of the trains (mean of pulses 3 to 5), the two wavelengths give similar photocurrents, 600 nA at 365 nm and 611 nA at 455 nm, which corresponds to an external quantum efficiency of about 15 % if the whole area is illuminated. The two wavelengths then produce opposite changes: the photocurrent increases by 38 % under 365 nm (potentiation), with most of the change happening within the first 20 pulses, while it decreases by 22 % under 455 nm (depression). The depression becomes stronger when the flux is increased by a factor of 4.7, reaching a decrease of 30 % (Figure 1c). Rather than requiring different electrical polarities to achieve potentiation and depression as it is done in most artificial synapses [30], the Se/ZnMgO photodiodes achieve bidirectional control through the spectrum of only the optical stimulus.

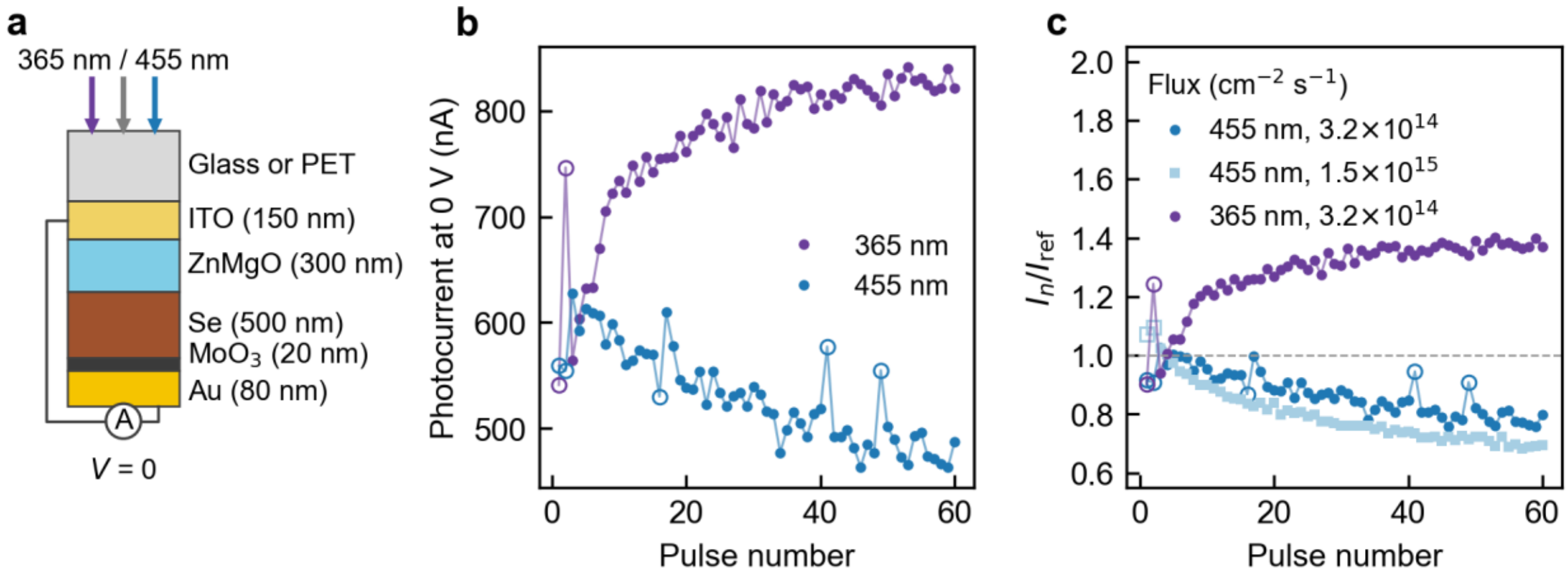

**Figure 1. Device architecture and bidirectional plasticity at zero bias.** (a) Schematic of the Se/ZnMgO photodiode; the photocurrent is recorded in short circuit. (b) Photocurrent of each pulse at zero bias for trains of 60 pulses of 5 ms at 20 Hz, under 365 nm and 455 nm light at the same photon flux ($3.2 \times 10^{14}$ $cm^{-2}$ $s^{-1}$). (c) Same data normalised to the mean of pulses 3 to 5 ($I_{ref}$), together with 455 nm pulses at $1.5 \times 10^{15}$ $cm^{-2}$ $s^{-1}$. Open symbols: points excluded by the processing rule, including the first two pulses of each train (see Experimental section).

Figure 2 shows how the frequency and the duration of the pulses change the response, at the maximum flux of each LED. Under 365 nm light, the photocurrent saturates within about ten pulses whatever the frequency (5 to 20 Hz) and the pulse duration (10 to 45 ms), and the plateau increases slightly with the photon dose, from about 906 to 963 nA between 5 and 20 Hz (Figure 2a,c). Under 455 nm light, the depression is slower and extends over the whole train, and it depends more strongly on the pulse parameters: higher frequencies give a faster and deeper depression, the photocurrent at the end of the train going from about 474 nA at 5 Hz to 402 nA at 20 Hz (Figure 2b), and longer pulses have the same effect, the photocurrent after 60 pulses decreasing from about 434 to 399 nA between 25 and 45 ms (Figure 2d). Since the runs of a series were acquired a few minutes apart, it is possible that the first pulse of each run partly still partly carries the state left by the previous one, and we therefore compare the absolute photocurrents rather than the normalised ones. These results show that the asymmetry between a fast, saturating potentiation and a slower, more gradual depression is a strong feature of the device, and that the rate of plasticity can be tuned through the number of photon delivered per pulse.

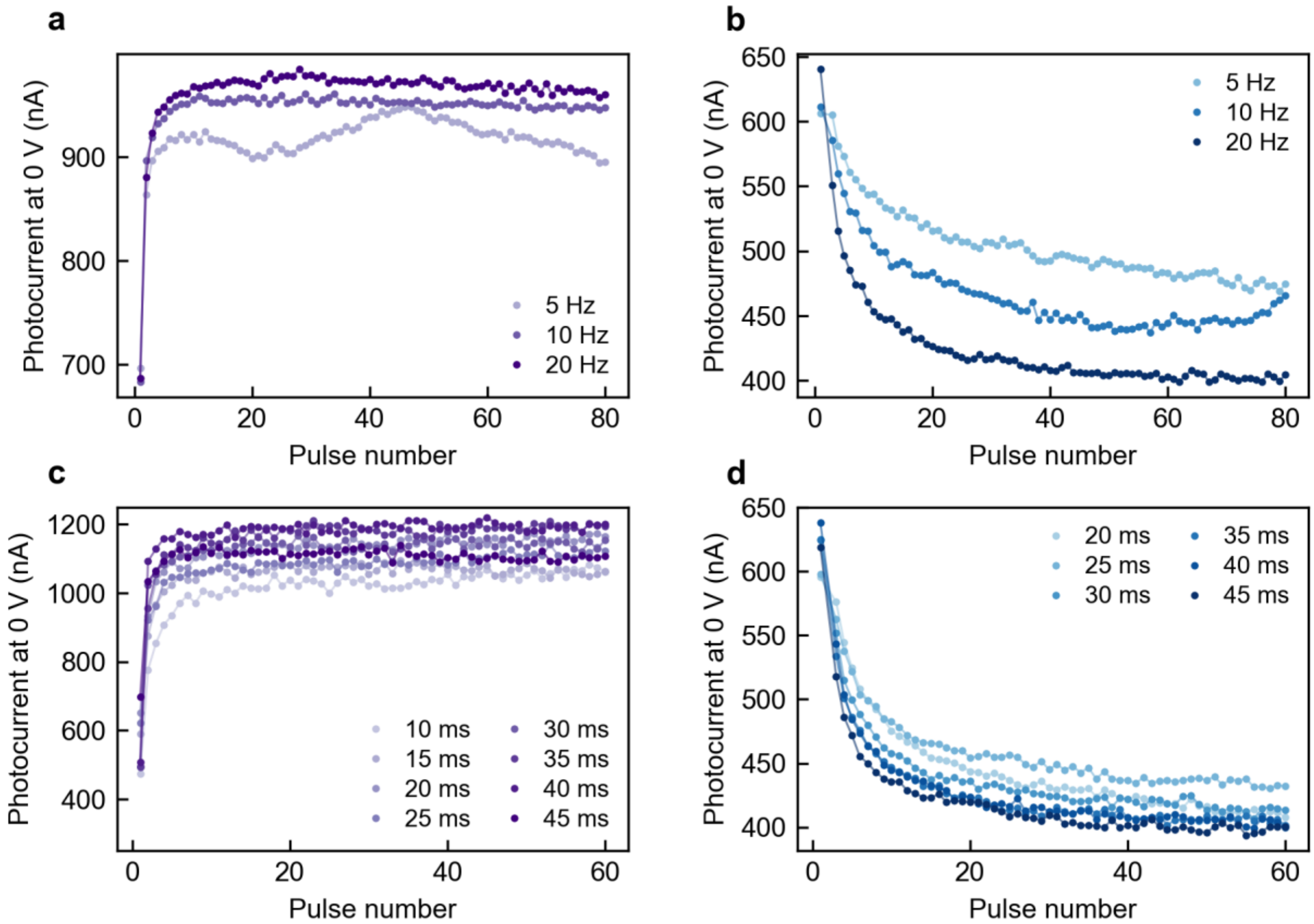

**Figure 2. Tuning the self-powered plasticity with the photon dose.** Photocurrent at zero bias as a function of pulse number. (a,b) Effect of the pulse frequency at fixed 40 ms pulse duration (80 pulses) for (a) 365 nm at $3.1 \times 10^{14}$ cm$^{-2}$ s$^{-1}$ and (b) 455 nm at $1.5 \times 10^{15}$ cm$^{-2}$ s$^{-1}$. (c,d) Effect of the pulse duration at 20 Hz (60 pulses) for (c) 365 nm and (d) 455 nm, at the same fluxes.

### 3.2. Origin of the wavelength selectivity

In order to understand what decides the sign of the change, we look at the effect of illumination over a wider spectral range. Figure 3a shows the change of the dark current of a batch B device after successive exposures at wavelengths from 1450 to 365 nm. After the exposures from 1450 to 505 nm, the dark current at 1 V changes by less than about 20 %, which is of the same order as the drift of the dark current observed in this measurement series without illumination, and we do not consider these small changes as significant enough to be meaningful. The dark current then increases by 60 % after 455 nm and by 208 % after 365 nm, which is well above this drift, so that the potentiation sets in at a photon energy between 2.45 (505 nm) and 2.73 eV (455 nm) for this batch. The exposures were made in order of increasing photon energy, so that the history of the device and the wavelength are not independent in this experiment. The effect of low-energy photons is however seen clearly on the potentiated state (Figure 3b): after the 365 nm exposure, 50 min in the dark only bring the dark current from 208 % to 144 %, while 40 min of 1450 nm light then bring it to 66 % and further exposures back to −6 %, which shows that low-energy photons actively erase the potentiated state rather than simply leaving it to relax.

This picture also helps to reconcile the two batches. In batch B, where potentiation sets in between 2.45 and 2.73 eV, 455 nm light (2.73 eV) is above the crossover and potentiates, both the dark conductance (Figure 3a) and the photocurrent at zero bias (Figure 5a,b). The crossover window of batch B is however obtained from the dark current, and for the photocurrent at zero bias only the sign of the change at 455 nm was measured, so that both observations reflect the same stored interface charge but their crossovers do not need to exactly coincide. IN batch A, 455 nm light depresses while 365 nm light potentiates, so that the crossover lies between 2.73 and 3.40 eV. In both batches, the shortest wavelengths potentiate and low-energy photons depress or erase, and in the end only the position of the crossover does change from one batch to the other, which we attribute to differences in the defect population of the interface between fabrication runs. The property of the Se/ZnMgO junction is therefore the competition between a potentiating and a depressing process whose balance depends on the photon energy, the crossover between the two being set by the defect population of the interface. Therefore, the batch-to-batch shift of the crossover is less a lack of reproducibility of the wavelength selectivity, but it is rather a direct consequence of its defect origin, and it therefore suggests that the crossover can be better controlled by controlling the interface during fabrication.

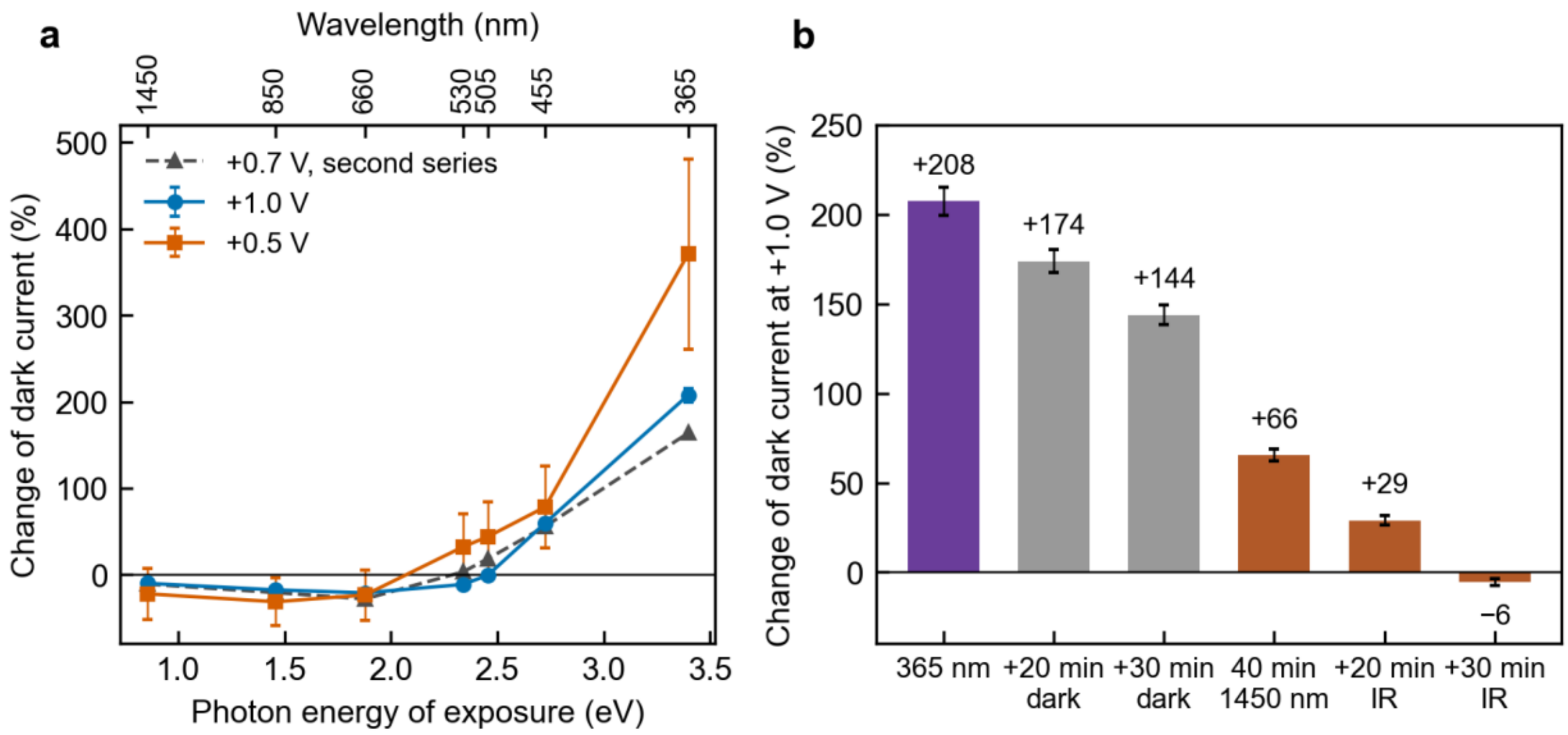

**Figure 3. Spectrally resolved light soaking (batch B).** (a) Change of the dark current with respect to the initial dark state after successive 5 min exposures at the indicated wavelengths, measured at 1.0 and 0.5 V (current-voltage sweeps) and at 0.7 V (second series); error bars are the standard error over a ±0.03 V window. Exposures were made in order of increasing photon energy; changes below about 20 % are of the order of the drift of the dark current in this series and are not considered significant. (b) Evolution of the dark current at 1.0 V after the 365 nm exposure: rest in the dark, then exposures to 1450 nm light.

A possible explanation for the different effect of the two wavelengths would be a different absorption profile, for instance if the ultraviolet light was absorbed in the ZnMgO window, where holes generated close to the junction could be trapped, while the blue light reaches the Se. We therefore calculated the absorption of each layer of the actual stack with measured optical constants (Figure 4). We also study the influence of Mg fraction of the sputtered ZnMgO between $x = 0$ and $x = 0.38$. For pure ZnO, 81 % of the 365 nm light is absorbed in the window and only 0.3 % reaches the Se, which would make the photocurrent at 365 nm much smaller than at 455 nm; this is in contradiction with the similar photocurrents measured at equal flux on Figure 1b, and this case can be excluded. As soon as $x \geq$ 0.12, the absorption edge of the window moves above 3.6 eV and the window becomes transparent at both wavelengths: the Se absorbs between 55 and 72 % of the 365 nm light and about 68 % of the 455 nm light, in agreement with the measured ratio of the photocurrents at the beginning of the trains (0.98). Moreover, the absorption coefficient of Se is high at both wavelengths, the absorption depth being 10.5 nm at 365 nm and 23 nm at 455 nm, so that 59 % of the incident photons are absorbed within the first 50 nm of Se next to the Se/ZnMgO interface at both wavelengths (Figure 4c). The two wavelengths therefore generate nearly the same number of carriers in the same region of the device. Closer to the interface however, the shorter absorption depth at 365 nm makes a difference: 21 to 29 % of the incident photons are absorbed within the first 5 nm of Se, against 13 % at 455 nm, so that the density of carriers generated right at the interface is about twice higher in the ultraviolet. This difference can contribute to the opposite behaviour of the two wavelengths in batch A. It cannot explain however the crossover of batch B between 455 and 505 nm, since 455, 505 and 530 nm light deposit nearly the same fraction of photons within 5 nm of the interface (13, 12 and 10 %), and for this batch the photon energy itself must decide the sign of the change.

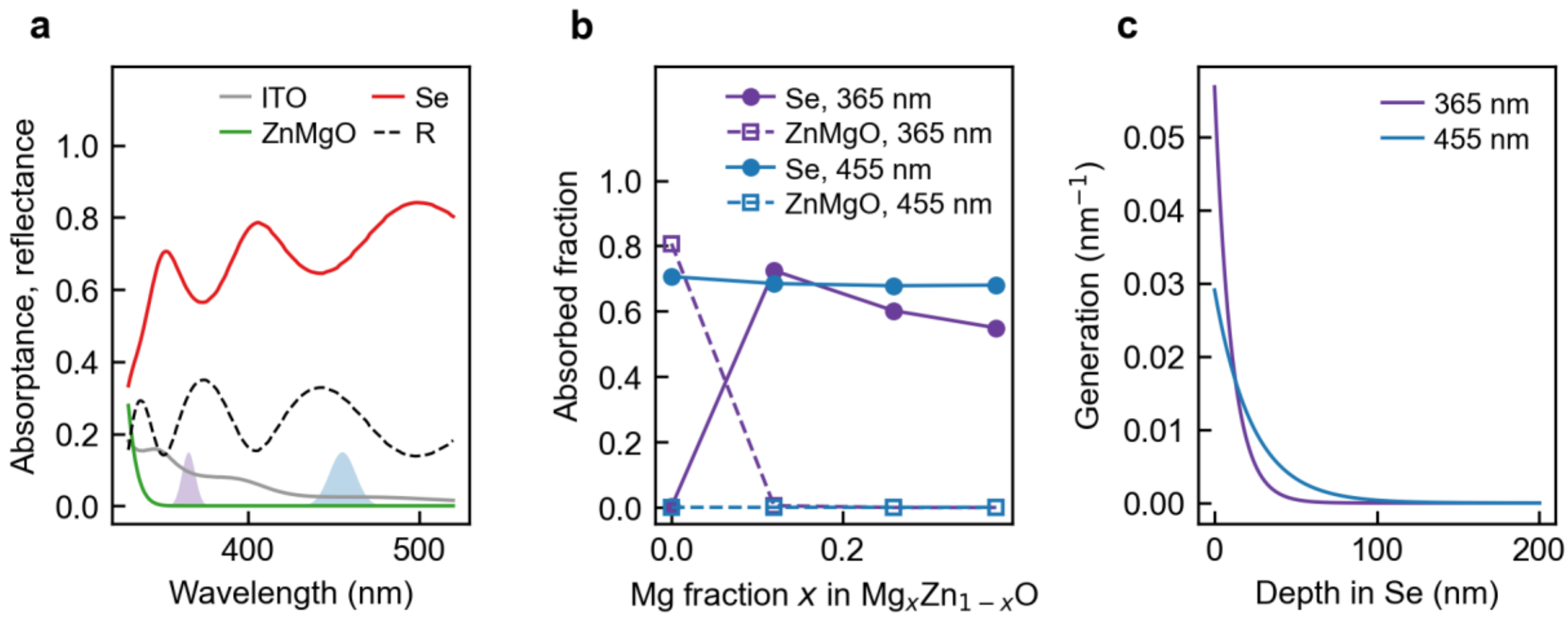

**Figure 4. Optical absorption in the device.** Transfer-matrix calculations with measured optical constants for all panels. (a) Absorptance of the ITO, ZnMgO and Se layers and reflectance of the stack for $Mg_{0.26}Zn_{0.74}O$; the shaded areas show the LED spectra. (b) Fraction of the LED light absorbed in the Se and in the ZnMgO window as a function of the Mg fraction of the window. (c) Generation profile in the Se layer for the two LEDs ($x$ = 0.26), per incident photon.

How the bidirectional plasticity depends on the quantity of photons, the crossover photon energy and the metastable character of the effect is consistent with charge-state transitions of defects near the Se/ZnMgO interface, where the carriers are generated (Figure 4c) and where metastable defects are known to accumulate in thin film solar cells [10]. In ZnO-based materials, the oxygen vacancy ($V_O$) is the most plausible candidate so far, and it was previously observed: it is a deep donor with a (2+/0) transition level about 1 eV below the conduction band, its charge-state transitions involve large lattice relaxations [31,32], and anion vacancies of this type have been identified as a source of persistent photoconductivity in II-VI and chalcopyrite semiconductors [29]. The direction of the charge transfer can be deduced from the electrical experiments presented below, independently of the defect identification. A reverse bias which extracts electrons from the interface region will erase the potentiated state and it brings the device below its initial state, while the forward read bias, which pushes electrons towards the interface, maintains the persistent conductance, but it disappears as soon as this bias is removed. Moreover, photons whose energy is too low to be absorbed in Se (1450 nm, below the Se band gap of about 1.95 eV) erase the potentiated state much faster than a rest in the dark (Figure 3b), which is consistent with an optical or photothermal emptying of trapped carriers rather than with a generation of carriers in the Se. Along with the drift-diffusion simulations presented below, these observations indicate that the potentiated state corresponds to interface states filled with electrons, and the depressed state to interface states emptied of their electrons. In terms of oxygen vacancies, potentiation corresponds to the capture of photogenerated electrons by ionised vacancies ($V_O^{2+} \rightarrow V_O^0$), which reduces the positive charge at the interface, and depression to the photoionisation of the neutral vacancies, which restores it. Under illumination, both processes will compete: electron capture is charged by the carriers generated in the Se next to the interface, while photoionisation is driven by the photons reaching the defects themselves. The crossover photon energy is thus the energy at which the two rates balance, and a wavelength close to the crossover is expected to show both components. The larger density of carriers generated at the interface by 365 nm light (Figure 4c) favours the capture, while the dependence of the crossover on the photon energy in batch B indicates that the photoionisation and capture rates also depend on the photon energy itself. The low external quantum efficiency of the devices (about 15 %, and it is decreasing at higher flux) indicates that the collection of the photogenerated carriers is limited by the junction, which is the regime where a change of the interface charge has a direct effect on the photocurrent.

In order to check whether a change of interface charge of this kind can reproduce the measured changes of photocurrent, we performed drift-diffusion simulations of the devices with SCAPS, calibrated to the measured external quantum efficiency of 13 to 15 % at both wavelengths (Table S1 and Figure S5). Reproducing a similar efficiency at 365 and 455 nm requires a moderately doped ZnMgO window ($5\times10^{16}$ to $1.5\times10^{17}$ cm$^{-3}$) and a small conduction band cliff at the interface, otherwise the carriers generated within the first nanometres of Se are lost and the ultraviolet response collapses. In the calibrated model, adding an additional positive charge at the interface decreases the photocurrent at both wavelengths, while adding a negative one increases it, and the effect is always stronger at 365 nm. This comes from the thin inverted layer on the Se side of the interface: a positive charge widens it and the holes generated inside it recombine before being collected, the carriers generated by 365 nm light, born within 10 nm of the interface, being the most affected. The potentiation measured at 365 nm (×1.38) is reproduced by a decrease of the positive interface charge of 2 to $7\times10^{10}\,q\,\text{cm}^{-2}$, and the depression measured at 455 nm (×0.78) by an increase of 4 to $13\times10^{10}\,q\,\text{cm}^{-2}$, depending on the window parameters. The two wavelengths therefore drive the interface charge in opposite directions, and ultraviolet light reduces the positive charge as expected from electron capture and blue light increases it as expected from photoionisation, which is consistent with the direction that we observed in the electrical experiments. The charge changes involved are small, of the order of $10^{10}$ to $10^{11}$ cm$^{-2}$, which is compatible with a minor fraction of the defects of a sputtered oxide interface changing charge state. The model in SCAPS contains no leakage path, and the dark current is about three orders of magnitude below the measured one so that it cannot be used to discuss the dark conductance.

The identification of the vacancy as the active defect relies on the defect physics of the ZnO family and it is still just a hypothesis at this point. The crossover energy measured here (2.45 to 3.4 eV depending on the batch) characterises the balance of the two processes rather than a single transition, and for a direct identification, it would require spectroscopic measurements of the buried interface, and this is beyond the scope of this work. Alternative mechanisms for persistent photoconductance in ZnO, in particular the photodesorption of oxygen from the surface [33,34], are unlikely here since the interface is buried beneath 500 nm of Se and a Au contact, and since a surface process would not explain the depression by low-energy photons nor the electrical reset discussed below.

If the synaptic state is stored in the charge state of interface defects, an applied field should modify it. Figure 5a,b shows the photocurrent at zero bias of a batch B device during successive 10 s exposures at 455 nm, which potentiate this batch. Ten exposures increase the photocurrent by about 30 to 50 %. When the device is then left in the dark, the stored state relaxes slowly: 40 % of the change remains after 5 min and about 2 % after 20 min. When a reverse bias of −1 V is applied in the dark instead, only 8 % of the change remains after 5 min, and a second period under bias brings the photocurrent 38 % below its initial value, after which the device is potentiated again by a new train of exposures. A short application of the reverse bias (10 s) is not sufficient, which shows that the reset is a gradual (analogue) and field-driven process and not just the effect of the bias switching itself.

The same behaviour is observed on the persistent photoconductance read under a forward bias of 0.5 V (Figure 5c,d). After a train of 455 nm pulses, interrupting the read voltage for 1 s already erases most of the persistent conductance, and applying a reverse voltage during that second drives the conductance even further below its initial value. Even at 0 V the current first falls below its initial value and recovers within about 10 s once the read bias is back: but with respect to the read bias, removing it is already a reverse step, so that the built-in field alone empties part of the states which were filled under 0.5 V before the train, and these refill in the dark under the read bias. The undershoot is larger for more negative voltages, the residual fraction 20 s after the reset going from −0.2 at 0 V to −1.0 at −1.4 V. With reset pulses applied at 25 Hz for 5 s, the residual fraction measured 0.5 s after the reset decreases from 0.55 at −0.25 V to 0.31 at −1 V (Figure 5e). The two readout modes have one important difference however, because in the biased mode, the forward read bias is itself part of what holds the stored state, so that removing it already erases most of the persistent conductance, while at zero bias the stored state relaxes over minutes and a reverse bias is needed to erase it within a few minutes. In both readout modes though, the stored state is erased by a reverse bias, even more strongly when the voltage is more negative, and it can be pushed below the initial state, which corresponds to an electrical depression. This is again consistent with a population of charged interface defects controlled by the electric field across the junction, the reverse bias increasing the field and emptying the states that store the potentiation.

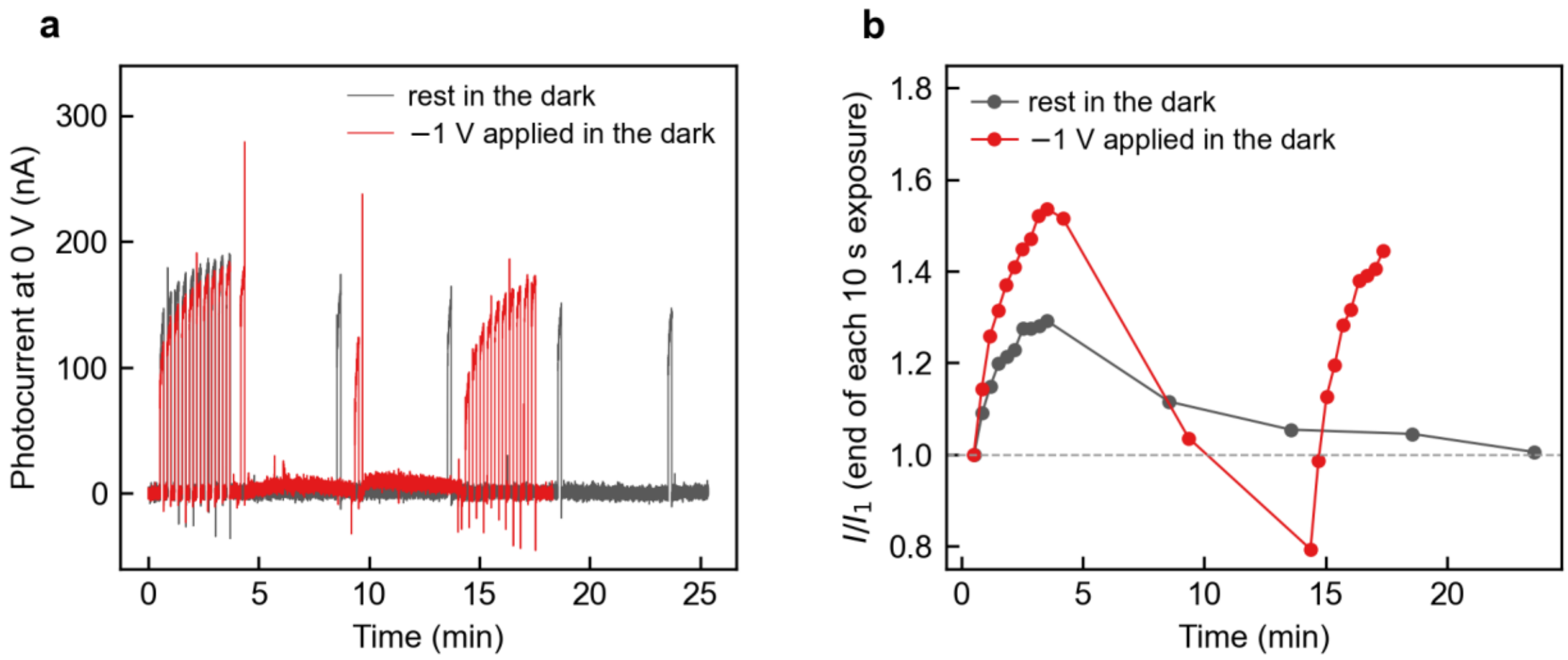

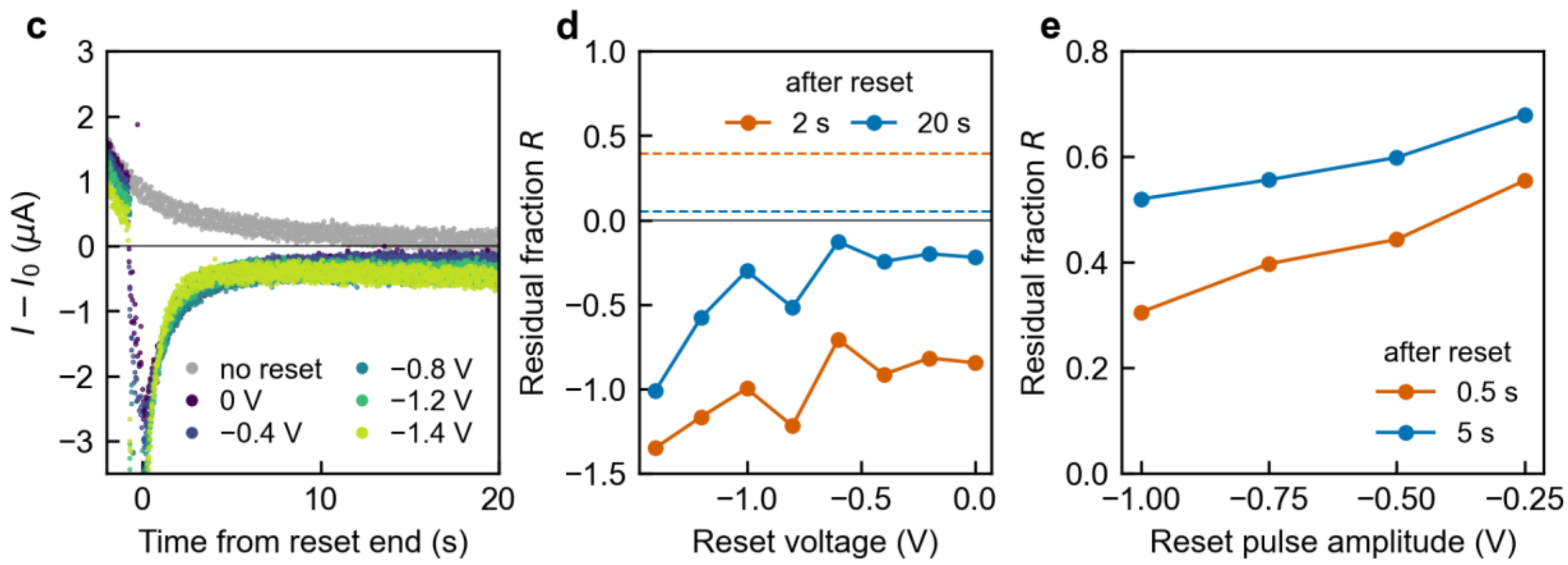

**Figure 5. Electrical reset of the stored state (batch B).** (a) Photocurrent at zero bias during 10 s exposures at 455 nm, with the device left in the dark (grey) or under −1 V in the dark (red) between the exposure series; within a series, the ten exposures are separated by 10 s in the dark at 0 V, the bias being applied only after the last one. (b) Photocurrent at the end of each exposure, normalised to the first one. (c) Current under a constant read bias of 0.5 V, interrupted only during the reset, after a 3 s train of 455 nm pulses (100 Hz, 5 ms) followed by a 1 s reset at the indicated voltage (0 V: read bias removed), aligned on the end of the reset, and without reset (grey). (d) Residual fraction $R = (I_{\mathrm{after}} - I_0)/(I_{\mathrm{before}} - I_0)$, 2 s and 20 s after the reset; dashed lines: without reset. (e) Residual fraction after reset pulses applied at 25 Hz for 5 s.

### 3.3. Synaptic functions and network simulation

An important property of adaptive memory is the ability to relearn previously acquired information faster than the first time, a phenomenon that is observed in biological networks and attributed to residual synaptic traces [35,36]. Figure 6a shows the conductance of a batch B device during two trains of 455 nm pulses at 10 Hz (5 s then 10 s) separated by 10 s in the dark, for incident power densities from 0.06 to 5.61 mW cm$^{-2}$. The conductance change after the first train increases with power, from 0.4 to 1.3 µS, and 40 to 50 % of it remains at the beginning of the second train (nearly all of it at 0.06 mW cm$^{-2}$). Thanks to this residual trace, the second train recovers the level reached at the end of the first one in 0.6 to 1.4 s instead of 5 s, which is 3.5 to 8 times faster, and it then continues to increase the conductance beyond that level. This speed-up does not show a clear dependence on the power within this range (Figure 6b), since the part of the change that has to be recovered also increases with the power. The rate at which the conductance increases at the beginning of the second train does however increase with the power, from 0.17 to 0.77 µS s$^{-1}$ between 0.3 and 5.61 mW cm$^{-2}$ (Figure 6c), and it increases faster than the rate of the first train: the ratio between the two goes from 0.57–0.64 below 1 mW cm$^{-2}$ to 0.76–1.11 above 2 mW cm$^{-2}$, so that the second train starts more slowly than the first one at low power, and as fast or faster at high power. The power densities were measured in sequence on the same device, from the lowest to the highest, so that this trend can also contain a contribution of the history of the device.

The same devices fabricated on PET allow to test the effect of mechanical bending, which we expected due to the fact that those devices have previously shown piezophototronic responses albeit with irreversible ITO damages [22]. Figure 6d shows the relative change of the current during and after a train of 1000 pulses at 160 Hz for the flat device, for three bending states, and for the flat device after the bending sequence. Two quantities have to be distinguished here. During the train, the current measured while the LED is on contains the photocurrent itself in addition to the conductance change, and we call it the response under illumination. Just after the last pulse, the current measured in the dark gives the stored change, which is the synaptic part of the response. The response under illumination is the largest in the flat state (28 %), it is reduced to 17 to 24 % for a displacement of 2.5 mm and to 12 % for the largest bending, and it goes back to 21 % when the device is flat again. Bending therefore modulates the response under illumination by up to a factor of two, partly reversibly, and the irreversible part (about 25 % after the whole sequence) is consistent with the cracking of the ITO electrode under bending which we reported for this device architecture and for ITO on PET [22,37]. The stored change behaves differently: it decreases from 12.3 % in the flat state to 7 to 8 % in all bending states and it stays at 6.9 % when the device is flat again, so that its reduction cannot be distinguished from the irreversible degradation. A piezopotential is generated in the ZnMgO layer under strain, as demonstrated on the photovoltaic performance of the same stack [22], and it is expected to contribute to the reversible part through the same field-controlled defect population as the piezopotential can act as an electrical reset; with a single device and one measurement per bending state, our data do not allow however to separate this contribution from the structural one. The pulse frequency dependence is preserved in all bending states (Figure 6e), so that the device keeps operating as a synapse on a flexible substrate.

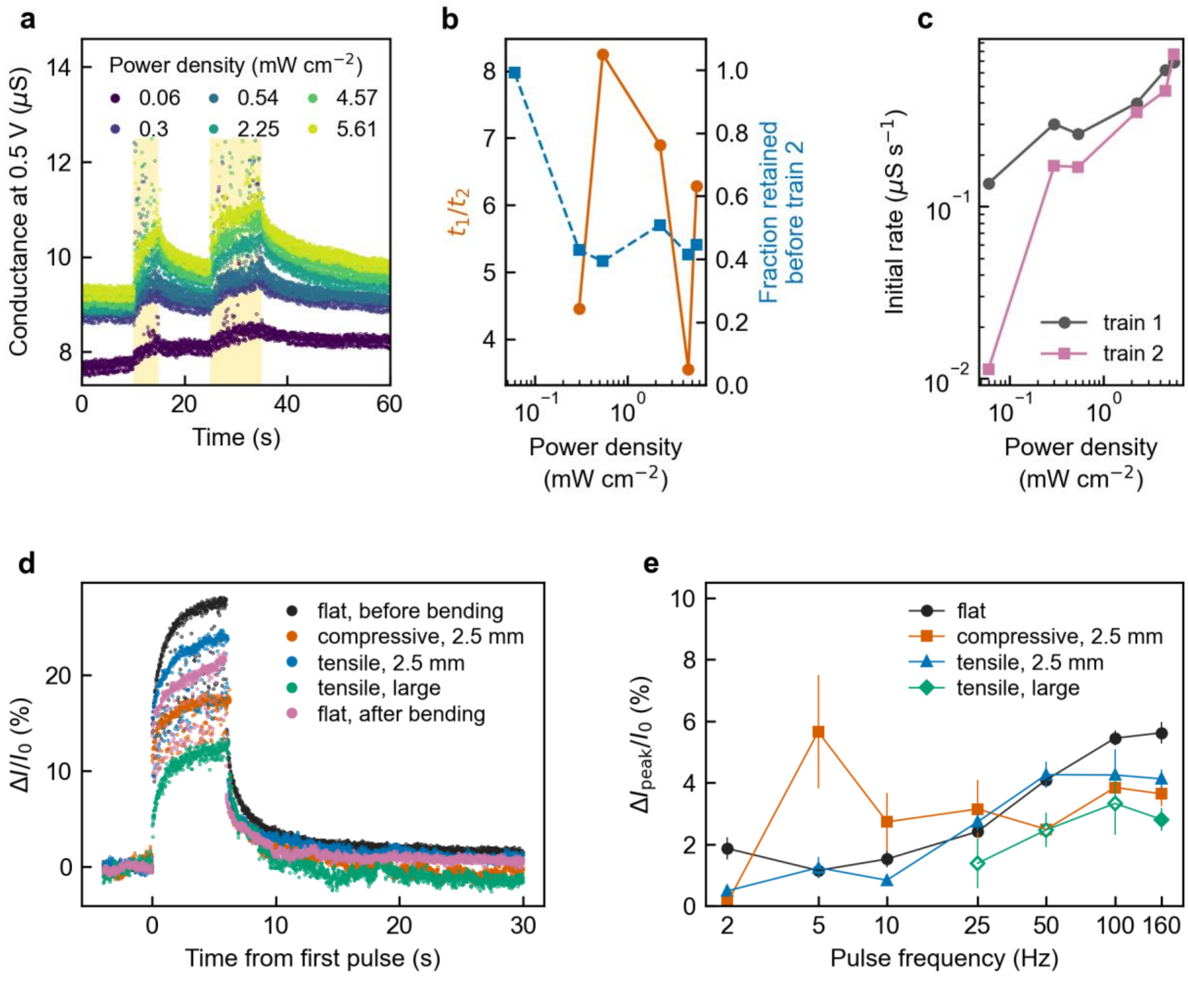

**Figure 6. Relearning and operation on a flexible substrate (batch B, 455 nm).** (a) Conductance at 0.5 V during two trains of pulses (10 Hz, 5 ms; shaded) for six incident power densities. (b) Ratio between the duration of the first train ($t_1$ = 5 s) and the time needed by the second train to reach the same level ($t_2$), and fraction of the first-train change retained before the second train. (c) Initial rate of the conductance increase during the first second of each train. (d) Relative change of the current under the read bias (about 0.7 V) during and after a train of 1000 pulses at 160 Hz for a device on PET, flat, under bending (displacements given in the legend), and flat after the bending sequence; during the train it includes the photocurrent in addition to the conductance change. (e) Relative change at the end of trains of 100 pulses as a function of the pulse frequency; open symbols: weak onset.

In order to assess the devices for in-sensor computing, we use the measured weight updates of Figure 1b in network simulations (Figure 7). The soft-bound model describes the potentiation well ($\alpha = 2.5 \pm 0.6$, $R^2 = 0.97$), with a strongly nonlinear update ($\nu = 0.14$), while the depression is close to linear over the measured range ($\nu = 0.63$); since the depression does not saturate within 60 pulses, its exponent is not well constrained, and only the measured range is used in the simulations. IN a 50 × 35 array with 10 % device-to-device variability, a grey-level image is written by giving each device a number of 365 nm pulses proportional to the grey level of its pixel (Figure 7d,f): the stored photocurrent map reproduces the image with a correlation coefficient of 0.88, the main loss coming from the saturation of the potentiation which compresses the brightest grey levels. A uniform exposure of the whole array to 58 pulses at 455 nm then removes 78 % of the contrast (Figure 7g), which illustrates the optical writing and erasing of an image in the array without any electrical signal. For handwritten digit recognition, a perceptron which weights are updated only through the measured potentiation and depression pulses of Figure 1b reaches 88.7 ± 1.0 % accuracy (ten seeds), against 97.5 % for floating-point weights, and 85 ± 2 % when the measured cycle-to-cycle scatter is included (Figure 7h). With the 10 ms pulses at maximum flux (Figure S2), the accuracy drops to 73.4 ± 9.3 %, and to 73.8 % with the scatter (the two values being within the spread), the large standard deviation over the seeds showing the fact that this device is at the limit of what the update rule can handle, since a single pulse moves the weight by up to 35 % of its range and the weights can no longer be adjusted in small steps. The loss compared with floating point therefore comes mainly from the size and the nonlinearity of the first updates rather than from the number of levels or the noise, and it decreases when the photon dose per pulse is reduced. This indicates that operating the potentiation further from saturation, with shorter or weaker 365 nm pulses, is the most direct route for improvement.

THese accuracies were obtained by assuming that 22 pulses span the whole window of the device, a choice which we did not optimise and whose effect is discussed in Section S6.

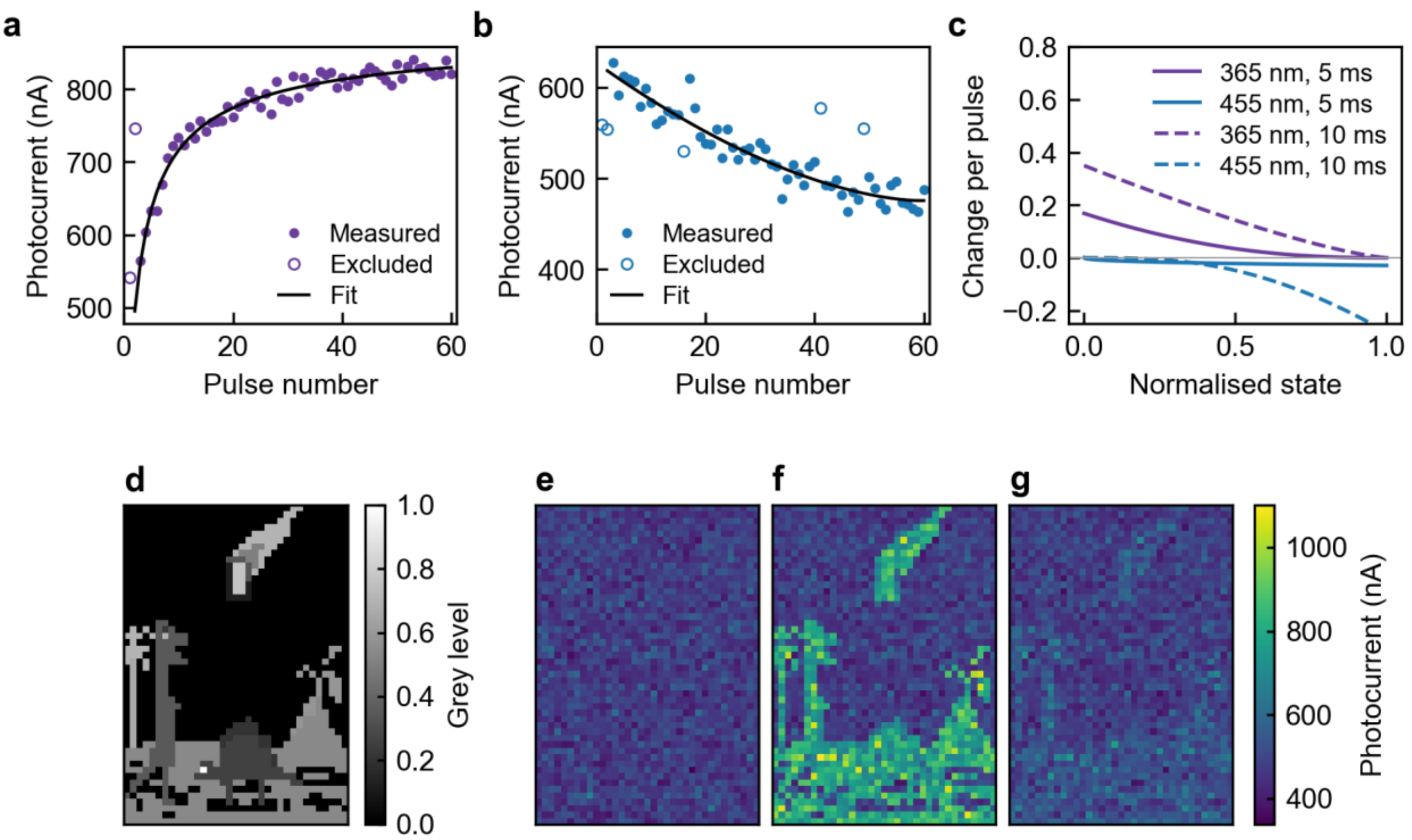

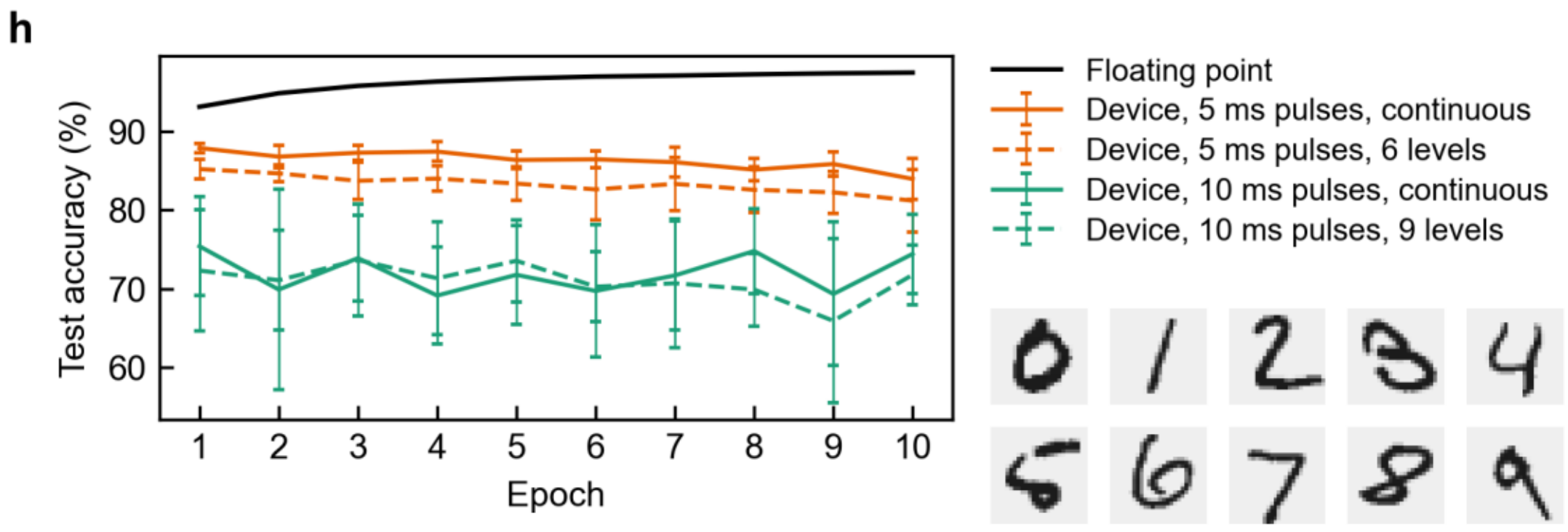

**Figure 7. Network simulation with the measured weight updates.** (a,b) Measured potentiation (365 nm) and depression (455 nm) of Figure 1b with the soft-bound model fitted from pulse 3 onwards; open symbols: excluded points. (c) Weight change per pulse as a function of the normalised state, for the trains of Figure 1b (5 ms pulses) and Figure S2 (10 ms pulses). (d) Grey-level image to be memorised (50 × 35 pixels). (e) Initial photocurrent map of a 50 × 35 array with 10 % device-to-device variability. (f) After writing with a number of 365 nm pulses proportional to the grey level (up to 58). (g) After a uniform exposure to 58 pulses at 455 nm. (h) Test accuracy of a 784-100-10 perceptron on MNIST with floating-point weights and with weights updated through the measured pulses, including the measured cycle-to-cycle scatter (mean and standard deviation over ten seeds), and examples of the handwritten digits of the MNIST test set; without scatter, the accuracy reaches 88.7 % (not shown).

Table 1 summarises the synaptic metrics. The number of distinguishable levels is estimated by dividing the excursion of the fitted train between pulse 3 and the last pulse by twice the pulse-to-pulse standard deviation $\sigma$, measured on the last 20 pulses after removal of the linear trend, and adding one. At equal flux and 5 ms pulses (Figure 1b), the potentiation extends over 270 nA with $\sigma$ = 9.3 nA, which gives 15 levels, while the depression at this low flux extends over 139 nA with $\sigma$ = 13.3 nA, which gives 6 levels. With 10 ms pulses and the maximum flux of each LED (Figure S2), the potentiation spans 226 nA ($\sigma$ = 13.1 nA, 9 levels) and the depression 471 nA ($\sigma$ = 12.8 nA, 19 levels). The incident optical energy is 56 to 70 nJ per pulse at equal flux and up to 0.5 µJ for the 10 ms pulses at maximum flux, for an area of 8 mm$^2$; this is larger than the values reported for micrometre-sized devices, and since the photocurrent density is the relevant quantity, it is expected to scale with the device area. No electrical energy is spent for writing or reading however.

**Table 1. Synaptic metrics of the Se/ZnMgO self-powered visual synapse (batch A, zero bias).** Relative changes are given with respect to the mean of pulses 3 to 5.

| Metric | Value | Source |
|---|---|---|
| Operating bias | 0 V (self-powered) | — |
| Device area | 8 mm$^2$ | — |
| Potentiation, equal flux (365 nm, 60 pulses) | ×1.38 | Fig. 1b |
| Depression, equal flux / 4.7 × flux (455 nm) | ×0.78 / ×0.70 | Fig. 1c |
| Distinguishable levels, pot. / dep. (5 / 10 ms pulses) | 15 / 6 and 9 / 19 | Fig. 1b, Fig. S2 |
| Nonlinearity $v$ (pot. / dep.) | 0.14 / 0.63 | Fig. 7 |
| External quantum efficiency (pulses 3 to 5) | ~15 % | Fig. 1b |
| Incident optical energy per pulse | 56–70 nJ | — |
| Electrical energy for writing and reading | 0 | — |
| Relaxation after potentiation (batch B) | 40 % left after 5 min, 2 % after 20 min | Fig. 5b |
| Relearning speed-up (batch B) | 3.5 to 8 | Fig. 6b |

| Metric | Value | Source |
|---|---|---|
| MNIST accuracy (device without noise (with noise) / floating point) | 88.7 % (85 %) / 97.5 % | Fig. 7 |

Table 2 places the present results in comparison with recent self-powered, or wavelength-sensitive, or strain-modulated optoelectronic synapses. The first group achieves zero-bias operation with a unidirectional optical response [11–14], optical erasing is being achieved in one case by reversing the direction of illumination rather than the wavelength [15]. The second group achieves all-optical bidirectional plasticity through changing the wavelength, but always under an applied read voltage [16–19]. The works that are the closest to the present work combine zero-bias reading with a wavelength-dependent sign of the response, either in a photoelectrochemical cell in an electrolyte [20] or through the competition between photovoltaic and photothermoelectric currents [21]. A third group demonstrates mechanical modulation of the plasticity, with a single stimulation wavelength and an electrical readout [38–42]. The Se/ZnMgO photodiodes combine zero-bias reading and wavelength-selective bidirectional plasticity within a single solid-state junction made of earth-abundant materials, with an electrical reset and operation on flexible substrates in addition.

**Table 2. Comparison with recent self-powered, wavelength-sensitive and strain-modulated optoelectronic synapses.** SP: self-powered (zero-bias) read; AOB: all-optical bidirectional plasticity; WSR: wavelength-selective sign reversal (one wavelength potentiates, another depresses); MM: mechanical modulation of the synaptic response.

| Material system | SP | AOB | WSR | MM | Ref. |
|---|---|---|---|---|---|
| Si nanocrystal-based devices | Yes | No | No | No | [11] |
| C8-BTBT/$F_{16}CuPc$ (organic) | Yes | No | No | No | [12] |
| $CsBi_3I_{10}$ (lead-free perovskite) | Yes | No | No | No | [13] |
| Triple-cation perovskite | Yes | No | No | No | [14] |
| $CsPbBr_3$ QDs/$C_3N_4$ | Yes | Yes | No[a] | No | [15] |
| VP/ZnO (photoelectrochemical)[b] | Yes | Yes | Yes (350/450 nm) | No | [20] |
| $SnO_2$/$CsBi_3I_{10}$–IR820[c] | Yes | Yes | Yes (650/820 nm) | No | [21] |
| $p^+$-Si/ZnO | No | Yes | Yes (405/633 nm) | No | [16] |
| ZnAlSnO/SnS | No | Yes | Yes (370/630 nm) | No | [17] |
| $Cu_2O$/$TiO_2$ | No | Yes | Yes (680/350 nm) | No | [18] |
| $PtSe_2$ (flexible) | No | Yes | Yes (405 vs 625/940 nm) | No[d] | [19] |
| ZnO/Ag nanowires | No | No | No | Yes | [38] |
| ZnO micro/nanowire | No | No | No | Yes | [39] |
| a-$Ga_2O_3$/ZnO | No | No | No | Yes | [40] |
| $Ga_2O_3$-based multimodal | No | No | No | Yes | [41] |
| SnSe/$SnSe_2$ | No | No | No | Yes | [42] |
| **Se/ZnMgO (this work)** | **Yes** | **Yes** | **Yes (365/455 nm)** | **Yes** | — |

[a] Optical erasing selected by the direction of illumination, not by the wavelength. [b] Two-electrode photoelectrochemical cell in aqueous electrolyte, not a solid-state device. [c] Sign reversal from competing photovoltaic and photothermoelectric effects. [d] Flexible and deformation-tolerant, without mechanical modulation of the plasticity.

# 4. Limitations

Several limitations should be however kept in mind. The qualitative behaviour is reproduced over the devices which we fabricated, but the crossover photon energy changes between the two fabrication batches. For an application, this means that the operating wavelengths have to be chosen on either side of the crossover of a given process, or that the interface has to be controlled well enough to fix the crossover, and a statistical study of the device-to-device and batch-to-batch dispersion remains to be done before array integration. We are not yet capable of fully controlling the defects at the interface and therefore the crossover. The retention, from minutes to tens of minutes, places the devices in the short-term plasticity regime, which is suitable for adaptive sensing and reservoir-type processing but not for long-term storage of weights. However, thin film solar cells with light soaking persisting over several days and possibly weeks have been observed, although rarely discussed since this kind of light soaking is often associated with low quality PV devices. Paired-pulse facilitation is not reported either, since it compares the first two pulses of a train, which the triggered acquisition does not resolve reliably; it would require a continuous sampling of the current. The network simulations deserve also some caution: the updates are open loop, the best operating point corresponds to a small number of pulses per update, and the case of the 10 ms pulses comes from two single-branch fits and not from a joint fit, with a starting state of the potentiation which is poorly constrained; the depression of this case was moreover measured at another flux than its potentiation, so that it has to be seen as an indication on the sensitivity to the update rule rather than as a prediction. Finally, mapping the crossover more finely in the same batch, as well as temperature-dependent measurements of the potentiation and depression rates, would allow a quantitative test of the proposed mechanism.

# 5. Conclusions

We have shown that Se/ZnMgO thin film photodiodes behave as self-powered visual synapses in which the photocurrent at zero bias writes and reads the synaptic weight, and in which the photon energy decides whether the weight increases or decreases. At equal photon flux, 365 nm pulses potentiate and 455 nm pulses depress the photocurrent, with up to 19 distinguishable levels per branch and no electrical energy spent. Spectrally resolved light soaking shows that short wavelengths potentiate while low-energy light erases, with a crossover photon energy which depends on the fabrication batch and is hard to control at this point, and optical simulations of the stack show that both wavelengths are absorbed in the first tens of nanometres of Se next to the heterojunction, the photon energy and the density of carriers generated at the interface deciding together the sign of the change. The voltage-dependent electrical reset, which can bring the device below its initial state, and the erasure by sub-gap light indicate that the stored state is a population of electrons trapped in interface defects and controlled by electric field, with the potentiation corresponding to their filling and the depression to their emptying. The devices show relearning, keep operating on flexible substrates, and a network built from the measured weight updates reaches 89 % accuracy on handwritten digit recognition (85 % with the measured noise). In conclusion, the metastable defect physics responsible for light soaking in thin film photovoltaics which is generally regarded as an instability to solve, can be used as a functional and spectrally controllable memory, and since such metastable defects are common to many thin film heterojunctions, other photovoltaic material systems may be used in the same way.

# CRediT authorship contribution statement

Conceptualisation, Z.J.L.K., M.K., K.J.T., and T.K.; methodology, Z.J.L.K. and T.K.; device fabrication, M.A.; investigation, Z.J.L.K., T.K., M.K., A.M., and M.A.K.; formal analysis, Z.J.L.K.; writing – original draft, Z.J.L.K.; writing – review & editing, M.K., S.G., S.K., and M.P.; funding acquisition, Z.J.L.K. and T.K.; supervision, Z.J.L.K. and T.K.

## Declaration of competing interest

The authors declare that they have no known competing financial interests or personal relationships that could have appeared to influence the work reported in this paper.

## Data availability

Raw measurement data for all figures, together with the processing scripts, will be deposited in the SOLIS project community on Zenodo (https://zenodo.org/communities/solis-project/) upon publication.

## Acknowledgements

This work has received funding from the European Union's Horizon Europe research and innovation programme under the Marie Skłodowska-Curie Staff Exchange project SOLIS (Grant Agreement No. 101183049). Financial support from the Spanish Ministry of Science, Innovation and Universities is acknowledged through the ULTRASYN project (PID2022-139445OA-I00, Proyectos de Generación de Conocimiento 2022). Z.J.L.K. thanks the Spanish Ministry of Science, Innovation and Universities for support through a Ramón y Cajal fellowship (RYC2021-033239-I). This research was partly supported by the JST FOREST Program (Grant Number JPMJFR246S) and JSPS KAKENHI (Grant Number 24K01422).

## Declaration of generative AI and AI-assisted technologies in the writing process

During the writing of this paper, the authors used Claude Opus 4.6 for part of the SYNAPSYS code. The experimental data were acquired by the authors, and the scientific content, interpretation and conclusions are from the authors. The authors reviewed take full responsibility for the content of the publication.

# Supplementary Information: Wavelength-selective bidirectional plasticity in self-powered Se/ZnMgO visual synapses

Zacharie Jehl Li-Kao[1,*], Miho Akasaka[2], Myeongok Kim[3], Kunal J. Tiwari[4], Antoine Moore[5], Mykhailo A. Koltsov[6], Sergio Giraldo[1], Saiki Kitagawa[2], Marcel Placidi[1], Taizo Kobayashi[2,**]

## S1. Stability of the read protocol

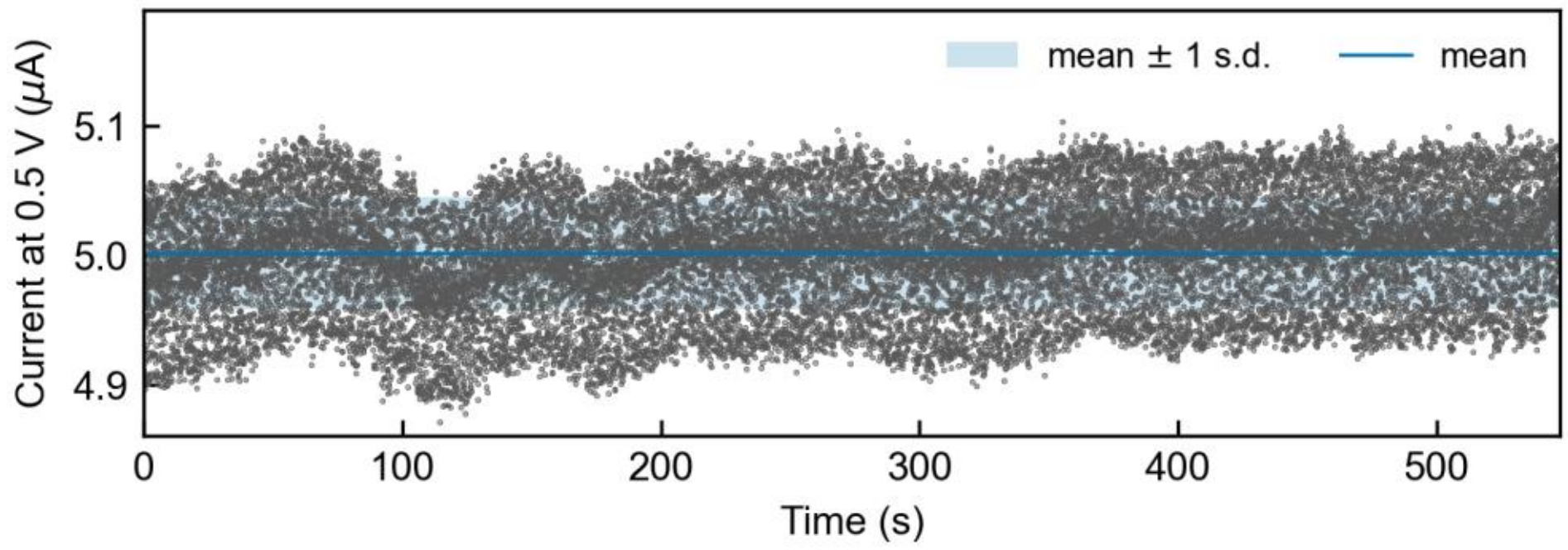

**Figure S1. Stability of the current under the read voltage alone (batch B).** Current at 0.5 V in the dark over 548 s (22805 samples); the band shows the mean ± one standard deviation. Mean 5.002 μA, standard deviation 0.044 μA (0.9 %), linear drift +0.06 % per minute, which is small with respect to the light-induced changes of Figures 5 and 6 of the main text.

## S2. Distinguishable levels

The number of distinguishable levels is estimated by dividing the excursion of the fitted train between pulse 3 and the last pulse by twice the pulse-to-pulse standard deviation $\sigma$, measured on the last 20 pulses after removal of the linear trend, and adding one. For the trains of Figure S2 (10 ms pulses at 20 Hz and the maximum flux of each LED), the potentiation spans 226 nA with $\sigma$ = 13.1 nA (9 levels) and the depression 471 nA with $\sigma$ = 12.8 nA (19 levels). For the trains of Figure 1b of the main text (5 ms pulses at equal flux), the same rule gives 15 levels in potentiation and 6 in depression.

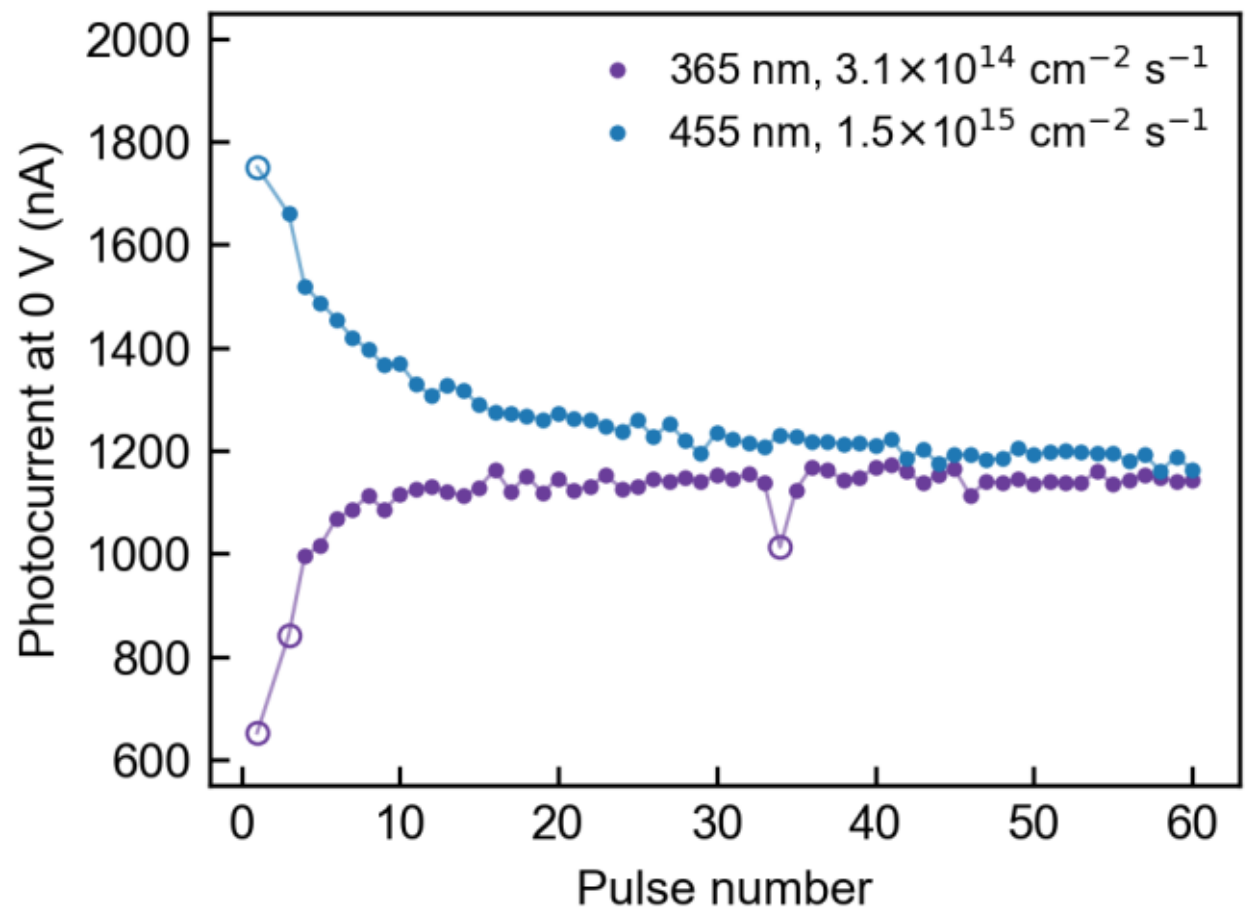

**Figure S2. Potentiation and depression with 10 ms pulses (batch A).** Photocurrent at zero bias, 60 pulses at 20 Hz, recorded three minutes apart. Open symbols: points excluded by the processing rule, including the first two pulses of each train.

## S3. Extraction of the photocurrent of each pulse

The multimeter records a block of 3 to 5 samples per trigger, one trigger per light pulse, and nothing between pulses. For every run, the value of a pulse is the mean of the samples of its block that exceed 50 % of the local bright level, defined as the median of the block maxima of the four neighbouring blocks; the excluded samples correspond to the rising edge of the LED or to a sample taken after the LED switched off. A block with fewer than half of its samples above this threshold is a spurious trigger and is discarded. A block whose number of retained samples differs from the most frequent number in the run caught a different part of the pulse, and a value further than four robust standard deviations (floored at 1 % of the signal) from the running median of five pulses is considered as an outlier; both are shown as open symbols and excluded from fits and plateau statistics. The first two pulses of each train are treated in the same way: at the start of a train, the timing of the trigger with respect to the LED pulse was not stable (compare the sample patterns of pulses 1 and 2 with the following ones in Figure S3) and the first pulse after a rest in the dark rises more slowly, so that their values cannot be compared with the rest of the train. Relative changes are given with respect to the mean of pulses 3 to 5. Figure S3 shows the raw samples and the extracted values for the runs of Figure 1b.

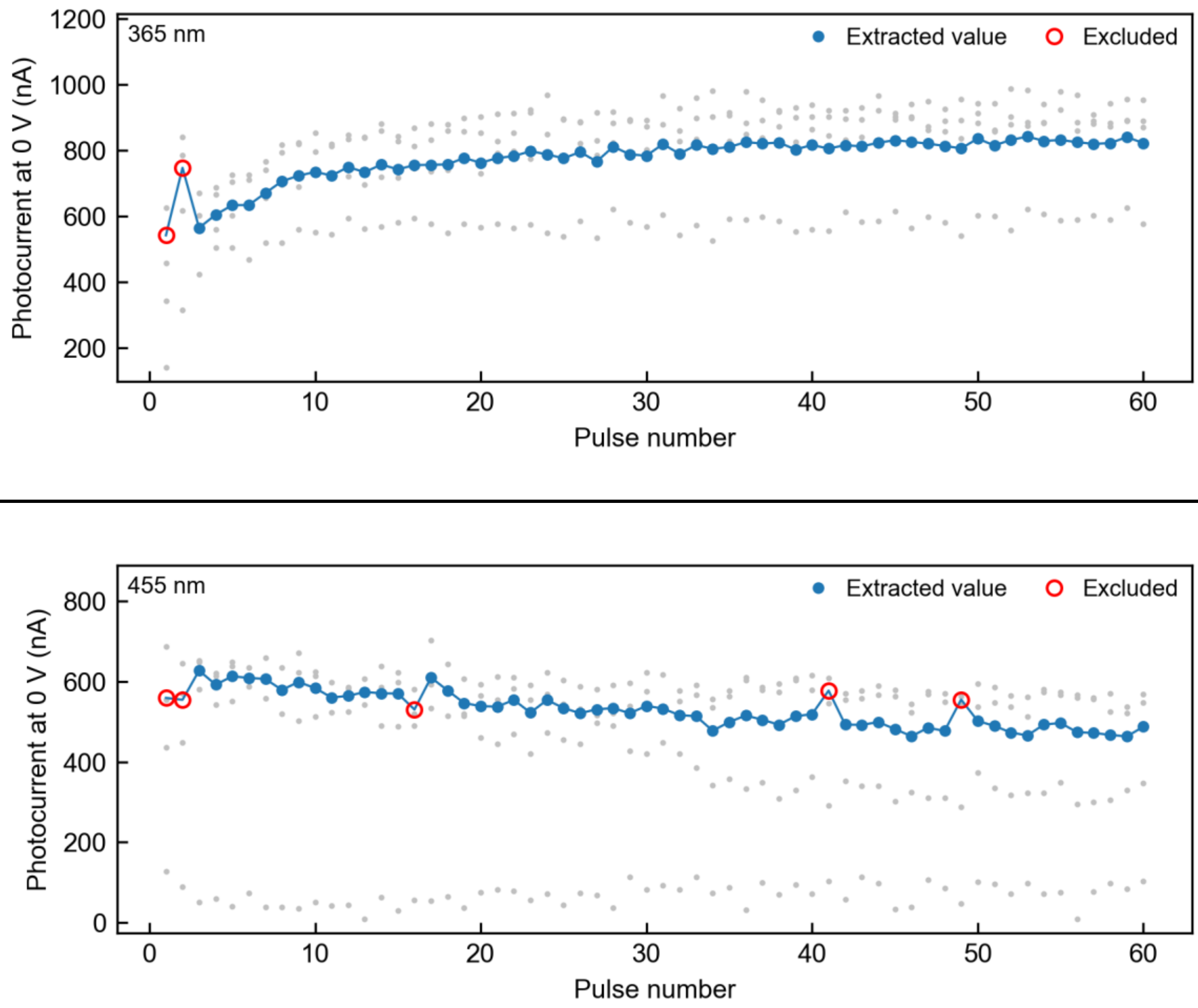

**Figure S3. Raw samples and extracted pulse values for the runs of Figure 1b.** Grey: all samples of each trigger block; blue: extracted value of each pulse; open red circles: excluded values.

## S4. Forward bias at zero-bias readout

Figure S4 shows a run of batch B in which a forward bias was applied in the dark between two series of exposures read at zero bias, instead of the reverse bias of Figure 5a,b of the main text.

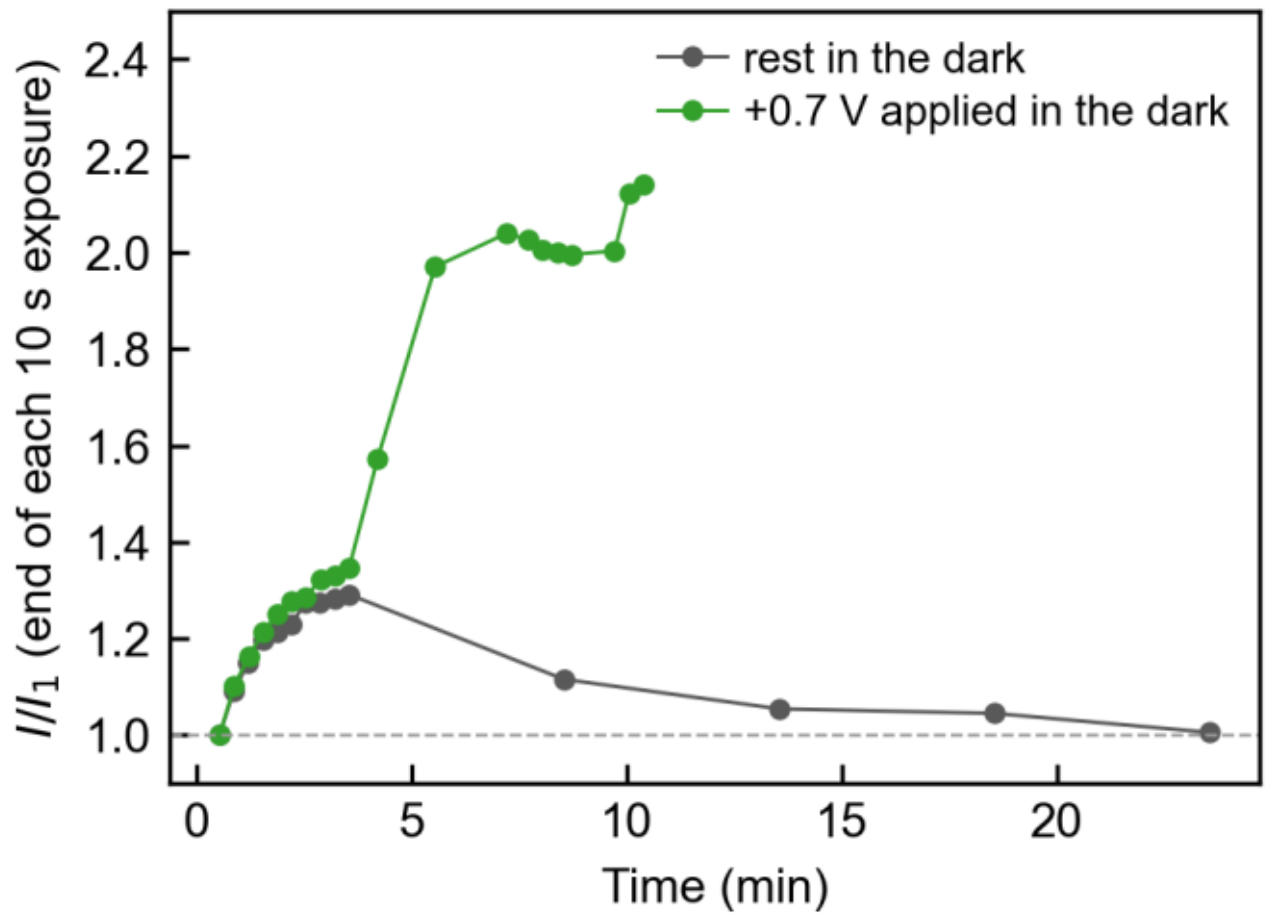

**Figure S4. Effect of a forward bias on the stored state (batch B).** Photocurrent at zero bias at the end of 10 s exposures at 455 nm, with the device left in the dark (grey) or under +0.7 V in the dark (green) after the first series of exposures. Instead of relaxing, the photocurrent increases further after the forward bias. In this run, a short period under negative bias was also applied according to the laboratory notes, so that the run is shown as a qualitative indication only.

## S5. Drift-diffusion simulations

The glass/ITO/ZnMgO/Se/$MoO_3$/Au stack was simulated with SCAPS-1D under monochromatic illumination at 365 and 455 nm ($3.2 \times 10^{14}$ photons $cm^{-2}\,s^{-1}$), with the parameters of Table S1. The baseline was calibrated to the measured external quantum efficiency of 13 to 15 % at both wavelengths, which gives 14.4 and 15.1 % (ZnMgO band gap of 3.6 eV) and 12.2 and 14.0 % (3.9 eV) at 365 and 455 nm. Amongst the parameters tested (Se mobility and lifetime, interface recombination, conduction band offset, back contact barrier, window doping), only the doping of the ZnMgO window allows to reproduce a similar efficiency at both wavelengths: a heavily doped window makes the Se next to the interface strongly n-type, and the holes generated there, mostly by 365 nm light, recombine before being collected. A change of interface charge $\Delta Q$ was introduced as a change of donor density in a 5 nm slab of ZnMgO at the interface (Figure S5); placing it as a sheet on the Se side gives the same trends. Since SCAPS does not include glass or interference, the absorptance of the Se in the model (0.91 at 365 nm, 0.98 at 455 nm) is higher than the values calculated with the full optical stack (Figure 4 of the main text), and the efficiencies are therefore comparable to the measured ones only approximately. The model contains no shunt or leakage path, and its dark current at +0.5 V (0.07 µA $cm^{-2}$) is about three orders of magnitude below the measured one, so that the dark conductance is not discussed with this model.

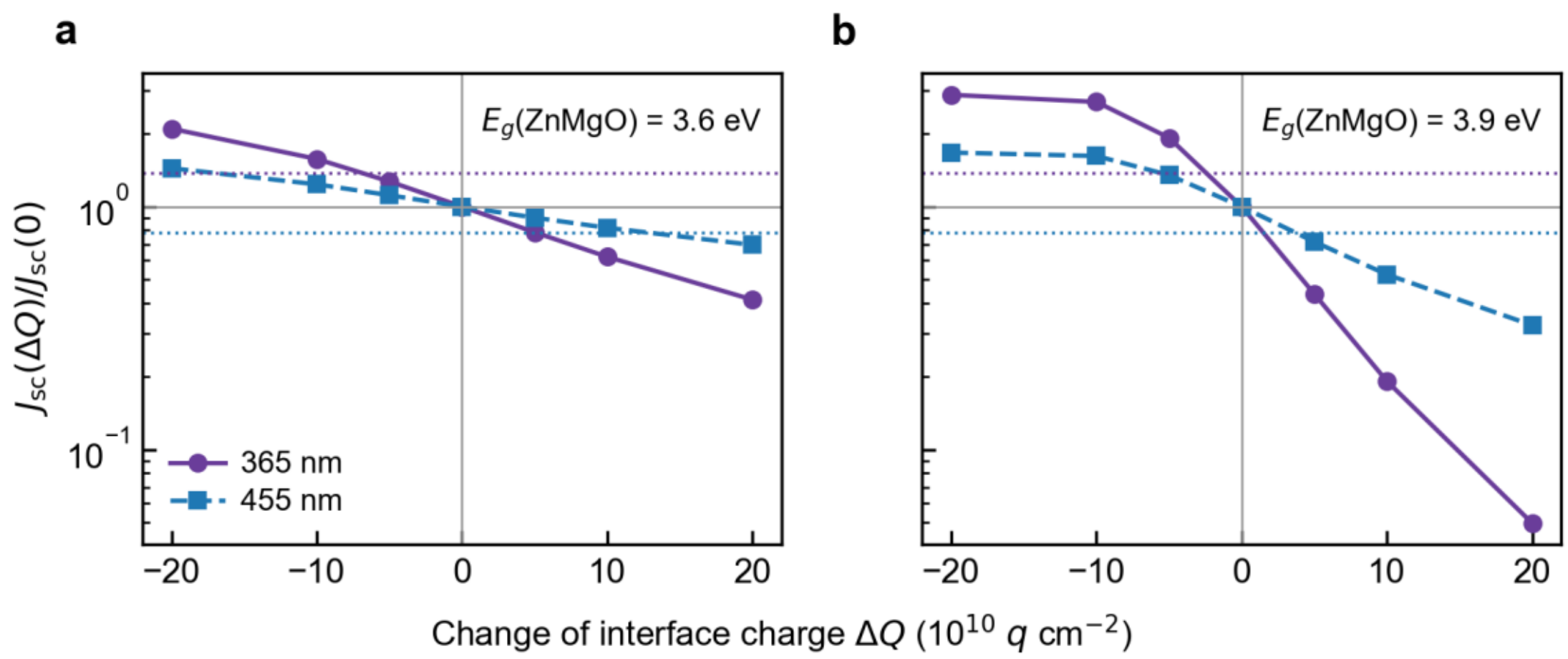

**Figure S5. Effect of a change of interface charge on the photocurrent at zero bias.** Ratio of the short-circuit current to its value without additional charge, at 365 nm and 455 nm, for a ZnMgO band gap of (a) 3.6 eV and (b) 3.9 eV; positive $\Delta Q$ corresponds to additional ionised donors. Dotted lines: measured ratios at equal flux (×1.38 at 365 nm, ×0.78 at 455 nm), with respect to the mean of pulses 3 to 5.

**Table S1. Parameters of the drift-diffusion simulations.**

| Parameter | Value | Source |
|---|---|---|
| Se band gap, electron affinity | 1.95 eV, 3.89 eV | [1,2] |
| Se permittivity, $N_c$, $N_v$ | 7.8, $8.7\times10^{18}$, $1.64\times10^{20}$ cm$^{-3}$ | [2] |
| Se acceptor density | $1\times10^{16}$ cm$^{-3}$ | [2] |
| Se electron and hole mobility | 0.46 cm$^2$ V$^{-1}$ s$^{-1}$ | [1] |
| Se bulk defects | $1\times10^{17}$ cm$^{-3}$, $\sigma = 1\times10^{-13}$ cm$^2$ (lifetime 10 ps) | calibrated |
| ZnMgO band gap | 3.6 and 3.9 eV | bracket (Mg fraction unknown) |
| ZnMgO donor density | $1.5\times10^{17}$ (3.6 eV), $5\times10^{16}$ cm$^{-3}$ (3.9 eV) | calibrated |
| ZnMgO electron affinity | 4.18 (3.6 eV), 3.97 eV (3.9 eV) | [1,3] |
| ZnMgO permittivity, $N_c$, $N_v$ | 8.0, $3\times10^{18}$, $1\times10^{19}$ cm$^{-3}$ | ZnO values |
| ZnMgO electron and hole mobility | 1.5, 0.1 cm$^2$ V$^{-1}$ s$^{-1}$ | [2] (electrons), ZnO value (holes) |
| Se/ZnMgO interface defects | $1\times10^{12}$ cm$^{-2}$, $\sigma = 1\times10^{-15}$ cm$^2$ | assumed (no effect on results) |
| $MoO_3$ | 20 nm, p-type $1\times10^{19}$ cm$^{-3}$, valence band aligned with Se | modelling choice |
| Contacts | flat-band ohmic, no series or shunt resistance | modelling choice |

# S6. Details of the network simulations

The network is a perceptron with 784 input, 100 hidden and 10 output units, with a ReLU activation in the hidden layer, a cross-entropy loss and a stochastic gradient descent with a learning rate of 0.1 and batches of 64 images, trained during 10 epochs on the 60 000 training images of MNIST and tested on its 10 000 test images. The initial weights are drawn from a Gaussian distribution of variance $1/N_{\text{in}}$, $N_{\text{in}}$ being the number of inputs of the layer, and the biases are kept in floating point, since they are not stored in the devices.

Each weight is stored as a differential pair of devices, $W = w_{\max}(g^+ - g^-)$ with $w_{\max} = 4/\sqrt{N_{\text{in}}}$, where $g \in [0,1]$ is the state of one device normalised to the window of its own branch: from the first pulse of the potentiation train to $G_{\max}$, and from $G_{\min}$ to the first pulse of the depression train, so that all the states remain inside the measured excursion. Since the two photocurrents are measured at two different wavelengths, they are not on the same conductance scale, and this normalisation to the own window is the convention of NeuroSim. The effect of one pulse (5 or 10 ms followed by a rest up to the period of 50 ms) is tabulated on 1025 points of the state from the fitted soft-bound model, with an Euler step of 1 ms. The case of the 10 ms pulses is built from two single-branch fits of the trains of Figure S2, and the starting state of its potentiation is poorly constrained, since the third pulse of the train is flagged by the processing rule.

The updates are open loop. The gradient gives a weight change $\Delta W$, which is converted into a number of pulses $n = |\Delta W| Q / w_{\text{max}}$, rounded in a stochastic way (the integer part, plus one with a probability equal to the fractional part) and limited to 8 pulses per update. $Q$ is the number of nominal pulses per window of the update, equal to 22 in the main text. For a positive $\Delta W$ the $n$ potentiating pulses are given to $g^+$ and the $n$ depressing pulses to $g^-$, and the opposite for a negative $\Delta W$, the state of the devices being never read in order to correct the update. The number of pulses is therefore proportional to the weight change, and a value of $Q$ below 1 means that a pulse is applied to a device only with a probability proportional to this change.

The read noise is a Gaussian fluctuation, drawn again at each forward pass, added to each state, and the write noise is a Gaussian fluctuation added to the new state after each pulse. The standard deviation is the plateau scatter divided by the window of the branch, which gives 0.024 (potentiation) and 0.093 (depression) for the 5 ms pulses, and 0.038 and 0.016 for the 10 ms pulses. In the conditions with levels, the states are rounded on a uniform grid with the smallest number of levels of the two branches (Table 1 of the main text).

**Update size.** In order to test the role of $Q$, the network was trained with $Q$ = 22, 11, 5.5, 2.75, 1.4 and 0.7. For each device, the value giving the best accuracy on a validation set was kept: the network was trained on 50 000 training images and tested on the 10 000 others (three seeds, 5 epochs), and the accuracy with the selected value was then measured on the test images after a training on the 60 000 images. On Figure S6a we can see the test accuracy for all the values (five seeds), and Table S2 gives the numbers with ten seeds for the value of the main text and for the selected one. The closed-form device of Figure S6b is described by the NeuroSim curve $w(x) = (1 - e^{-x/\nu})/(1 - e^{-1/\nu})$, with 50 pulses per window and the same nonlinearity for potentiation and depression. The number of nominal pulses per window in the update is $Q = sP$, with $P = 50$ and a scale $s$ between 1/256 and 1 selected for each nonlinearity on 10 000 validation images, and the network was trained during 5 epochs on 50 000 images (three seeds), so that the ideal network of this panel is at 96.4 % and not at 97.5 %.

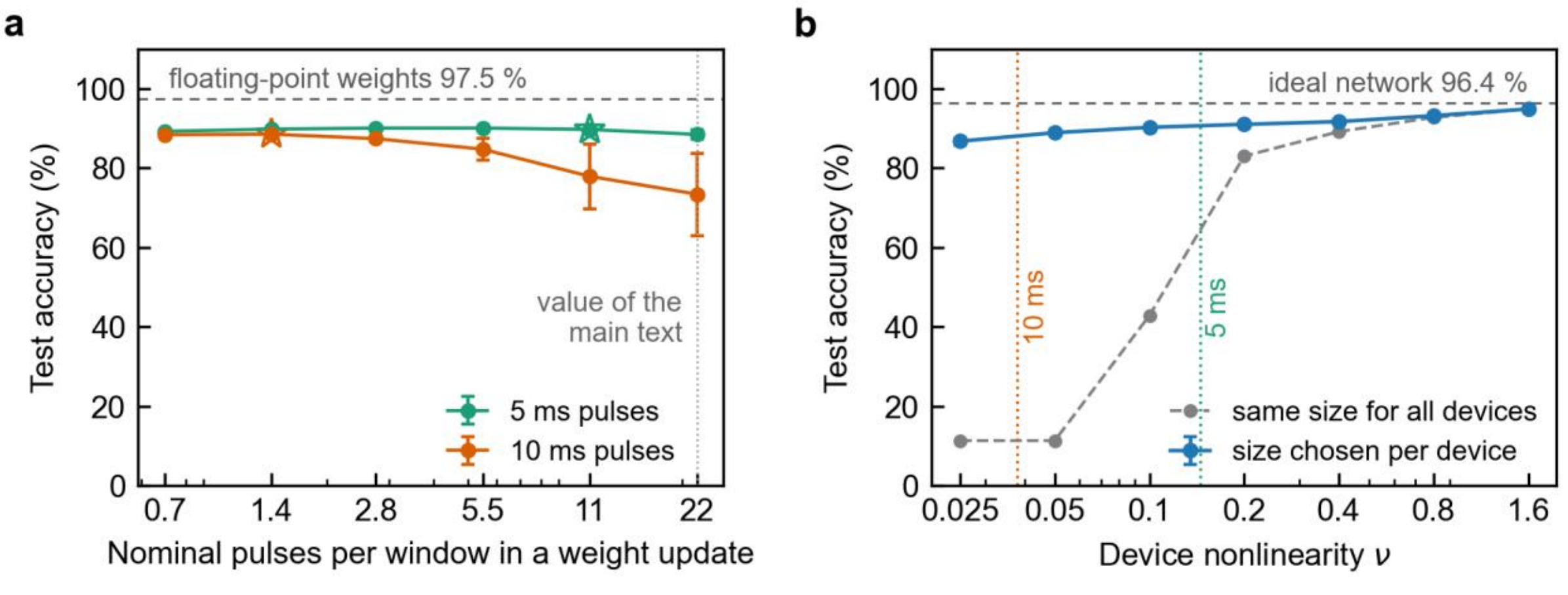

**Figure S6. Role of the number of pulses applied for a weight update.** (a) Test accuracy of the two devices of the main text against the number of nominal pulses per window in the update (mean and standard deviation over five seeds, 10 epochs); the value of the main text is 22, and the star marks the value selected on the validation set.

(b) Closed-form device described by its nonlinearity $v$ only, symmetric for potentiation and depression, with the update size chosen for each device or the same for all of them (50 000 training images, 5 epochs, three seeds, range as error bars); the dotted lines show the nonlinearity of the potentiation of the two devices.

**Table S2. Test accuracy for the number of pulses per window of the main text ($Q$ = 22) and for the value selected on the validation set** (continuous states, mean ± standard deviation over ten seeds, twenty for the write noise at $Q$ = 22, floating-point weights 97.5 %).

| Pulses | $Q$ | No noise (%) | Write noise (%) | Seeds |
|---|---|---|---|---|
| 5 ms | 22 | 88.7 ± 1.0 | 85.1 ± 2.3 | 10 / 20 |
| 5 ms | 11 (selected) | 89.8 ± 0.2 | 87.5 ± 0.8 | 10 |
| 10 ms | 22 | 73.4 ± 9.3 | 73.8 ± 5.1 | 10 / 20 |
| 10 ms | 1.4 (selected) | 88.4 ± 0.5 | 88.0 ± 0.5 | 10 |